\documentclass[a4paper,11pt]{article}
\usepackage{pos}
\usepackage{tikz}
\usetikzlibrary{decorations.markings,decorations.pathmorphing}
\usetikzlibrary{fit, calc}
\usetikzlibrary{angles,quotes} 
\usetikzlibrary{matrix,arrows.meta,calc} 
\usepackage{amsmath}
\usepackage{cancel} 
\usepackage[dvipsnames]{xcolor}
\usepackage{subcaption}
\usepackage{braket}
\usepackage{tensor}
\usepackage{etoolbox} 
\usepackage{graphicx}  
\usepackage{comment}

\usepackage{scalerel}   
\usepackage{stackengine} 
\stackMath

\newcommand{\CancelTo}[3][]{%
  \ifblank{#1}{}{%
    \renewcommand{\CancelColor}{#1}%
  }
  \cancelto{#2}{#3}%
}

\renewcommand{\eqref}[1]{\hyperref[#1]{Eq. \textup{(\ref{#1})}}}
\newcommand{\secref}[1]{\hyperref[#1]{Section}~\textup{\ref{#1}}}
\newcommand{\figref}[1]{\hyperref[#1]{Fig. \textup{\ref{#1}}}}
\makeatletter
\newcommand{\reallywidetilde}[1]{\mathpalette\reallywidetilde@{#1}}
\newcommand{\reallywidetilde@}[2]{%
  \begingroup
  \setbox0=\hbox{$\m@th#1#2$}%
  \mathop{#1#2}\limits^{%
    \smash{\raisebox{-0.45\ht0}{%
      \scalebox{1}[0.88]{
        \resizebox{\wd0}{!}{$\m@th#1\sim$}%
      }%
    }}%
  }%
  \endgroup
}
\makeatother

\title{Subleading soft dressings for QED scattering states}

\author*[a]{Stavros Christodoulou}
\author[a]{Nicolaos Toumbas}

\affiliation[a]{Department of Physics, University of Cyprus\\
  Nicosia 1678, Cyprus}

\emailAdd{christodoulou.stavros@ucy.ac.cy}
\emailAdd{nick@ucy.ac.cy}

\abstract{
    We study soft emission in QED during scattering of Faddeev-Kulish dressed states. The incoming and outgoing charged particles are accompanied by coherent clouds of soft photons with energies below a characteristic infrared scale $E_d$. We focus on explicit processes that allow the dependence of the soft factors on the hard particles’ momenta and total angular momenta to be displayed clearly. We argue that the dressings remove the infrared divergences in hard amplitudes order by order in perturbation theory, effectively regulating the contributions from virtual soft photons at the scale $E_d$. Essentially, the Faddeev–Kulish hard amplitudes become equivalent to the infrared-finite part of the corresponding Fock-basis amplitudes. Finally, tree-level soft-photon emission is found to be suppressed once the dressings are extended to subleading order in the soft-momentum expansion, as prescribed in recent work by Choi and Akhoury.
}

\FullConference{
}

\begin{document}
\maketitle

\section{Introduction}
    \label{sec:intro}
    The S-matrix elements between conventional Fock states in gauge theories and gravity suffer from infrared divergences \cite{Yennie:1961ad,Weinberg:1965nx,Gell-Mann:1954wra,Bloch:1937pw,Low:1954kd,Low:1958sn,Burnett:1967km,Dollard:1964jmp,Chung:1965zza,Greco:1967zza,Kibble:1968sfb,Kibble:1968lka,Kibble:1968npb,Kibble:1968oug,Kulish:1970ut,Strominger:2017zoo}. The divergences, which appear at any finite order in perturbation theory for generic Fock-basis states, stem from the fact that interactions mediated by massless gauge bosons and gravitons 
are not completely turned off at large times. Recent developments have revealed deep connections between these infrared phenomena and asymptotic gauge symmetries. Specifically, scattering processes are constrained by an infinite set of conservation laws associated with large gauge transformations
\cite{Lysov:2014csa,He:2014cra,Kapec:2015ena,Campiglia:2015qka,Mohd:2014oja,Gabai:2016kuf,Strominger:2017zoo,Kapec:2017tkm,Bondi:1962px,Sachs:1962wk,Strominger:2013lka,Campiglia:2016hvg,Carney:2017jut}. 

QED
provides a simple and natural and testing ground for exploring the deep infrared structure, many features of which are shared by non-abelian gauge and gravitational theories alike. Infrared divergences in QED can be classified into two categories, those associated with virtual photons \cite{Yennie:1961ad,Weinberg:1965nx} and those associated with real soft-photon emission \cite{Bloch:1937pw,Low:1954kd,Low:1958sn,Burnett:1967km}. 
These two types of infrared divergences in scattering amplitudes suggest the need to move away from the conventional Fock basis of asymptotic states, used to compute S-matrix elements, and seek a refined formulation in terms of a basis consistent with gauge invariance, under the group of large gauge transformations whose gauge parameters tend to angle dependent constants at infinity  \cite{Kulish:1970ut,Kapec:2017tkm,Campiglia:2015qka}.

\begin{figure}[!t] 
    \centering
    \begin{minipage}[c]{\linewidth}
    \vspace{0.5em}
    \centering
        \begin{minipage}[c]{0.45\textwidth}
            \centering
            \resizebox{0.875\linewidth}{!}{\begin{tikzpicture}
[line width=1pt, 
  fermion/.style={thick},  
  boson/.style={decorate, decoration={snake}, draw=black},
  softboson/.style={decorate, decoration={snake}, draw=blue},
  halffermion/.style={thick, ->}, scale=1.5]

  \filldraw[fill=gray!20, draw=black] (0,0) circle (0.5);

  \draw[fermion] (-2,1) -- (-0.5,0.25);
    \node at (-2.25,1) {$p_1$};
  \draw[fermion] (-2,0) -- (-0.5,0.0);
    \node at (-2.25,0) {$p_2$};
  \draw[fermion] (-2,-1) -- (-0.5,-0.25);
    \node at (-2.25,-1) {$p_3$};
  
  \draw[halffermion] (-2,1) -- (-1.25,0.625);
  \draw[halffermion] (-2,0) -- (-1.25,0);
  \draw[halffermion] (-2,-1) -- (-1.25,-0.625);

  \draw[fermion] (0.5,0.25) -- (2,1);
    \draw[halffermion] (0.5,0.25) -- (1.25,0.625);
    \node at (2.25,1) {$q_1$};
  \draw[fermion] (0.5,0.0) -- (2,0);
    \draw[halffermion] (0.5,0.0) -- (1.25,0);
    \node at (2.25,0) {$q_2$};
  \draw[fermion] (0.5,-0.25) -- (2,-1);
    \draw[halffermion] (0.5,-0.25) -- (1.25,-0.625);
    \node at (2.25,-1) {$q_3$};


\end{tikzpicture}}
            \subcaption{}
        \end{minipage}
        \begin{minipage}[c]{0.45\textwidth}
            \centering
            \resizebox{0.875\linewidth}{!}{\begin{tikzpicture}
[line width=1pt, 
  fermion/.style={thick},  
  boson/.style={decorate, decoration={snake}, draw=black},
  softboson/.style={decorate, decoration={snake}, draw=blue},
  halffermion/.style={thick, ->}, scale=1.5]

  \filldraw[fill=gray!20, draw=black] (0,0) circle (0.5);

  \draw[fermion] (-2,1) -- (-0.5,0.25);
    \node at (-2.25,1) {$p_1$};
  \draw[fermion] (-2,0) -- (-0.5,0.0);
    \node at (-2.25,0) {$p_2$};
  \draw[fermion] (-2,-1) -- (-0.5,-0.25);
    \node at (-2.25,-1) {$p_3$};
  
  \draw[halffermion] (-2,1) -- (-1.25,0.625);
  \draw[halffermion] (-2,0) -- (-1.25,0);
  \draw[halffermion] (-2,-1) -- (-1.25,-0.625);

  \draw[fermion] (0.5,0.25) -- (2,1);
    \draw[halffermion] (0.5,0.25) -- (1.25,0.625);
    \node at (2.25,1) {$q_1$};
  \draw[fermion] (0.5,0.0) -- (2,0);
    \draw[halffermion] (0.5,0.0) -- (1.25,0);
    \node at (2.25,0) {$q_2$};
  \draw[fermion] (0.5,-0.25) -- (2,-1);
    \draw[halffermion] (0.5,-0.25) -- (1.25,-0.625);
    \node at (2.25,-1) {$q_3$};

  
  \draw
    [softboson, decoration={snake, amplitude=2pt, 
      segment length=0.25cm}]
  (-1.85, 0)
    .. controls (-1.8, -0.15) and (-1.7, -0.3)
    .. (-0.975, -0.475);
  \draw
    [softboson, decoration={snake, amplitude=2pt, 
      segment length=0.2635cm}]
    (-1.625, 0.8125)
      .. controls  (-1.2, 1.0) and (1.2, 1.0)
      .. (1.625, 0.8125);
\end{tikzpicture}}
            \subcaption{}
        \end{minipage}
    \end{minipage}
    \caption{
        Feynman diagrams for a generic hard scattering process in QED, including a diagram with soft virtual photon insertions (right). The soft virtual  photon propagators are drawn in blue. Such diagrams give rise to infrared logarithmic divergences, which exponentiate and lead to the vanishing of purely hard amplitudes.
        }
    \label{fig:virtual_loops}
\end{figure}
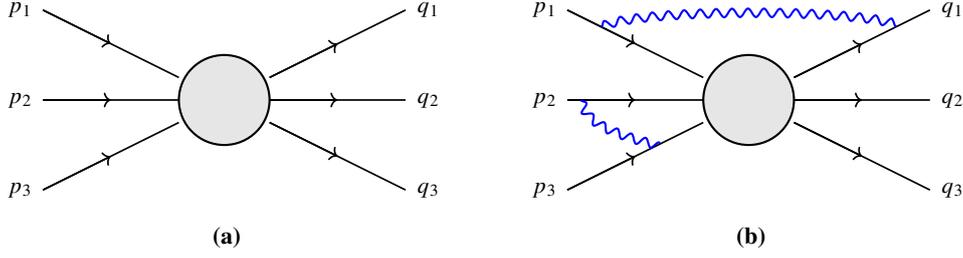

Virtual infrared divergences appear at the loop level
as logarithms of the infrared cutoff, which we denote by $\lambda$. When summed to all orders, these logarithmic divergences exponentiate, leading to the vanishing of generic hard scattering amplitudes in the $\lambda\to 0$ limit. Concretely, for a scattering process $\alpha\to\beta$, where $\alpha$ and $\beta$ denote the initial and final asymptotic hard states, it can be shown that \cite{Weinberg:1965nx}
\begin{equation}\label{eq:elastic_vanishing}
    S_{\beta\alpha}=
        e^{i\phi_{\beta\alpha}}~
        e^{\mathcal{B}_{\beta\alpha}~
            \ln(\lambda/\Lambda)}~
        S_{\beta\alpha}^{(\Lambda)}
\end{equation}
Here, $\phi_{\beta\alpha}$ is a phase, which does not contribute to the square of the amplitude, 
and $\mathcal{B}_{\beta\alpha}$ is a positive kinematical constant. See \eqref{eq:kinematical_factor} below for the explicit expression of $\mathcal{B}_{\beta\alpha}$. In addition, $\Lambda > \lambda$ (taken to be sufficiently smaller than the electron mass) is an infrared scale distinguishing soft from hard virtual photons, and $S_{\beta\alpha}^{(\Lambda)}$ is the infrared finite part of the S-matrix element, 
stripped from the contributions of the soft virtual photons. Since $\mathcal{B}_{\beta\alpha}$ is positive and $\Lambda > \lambda$, the S-matrix element vanishes in the limit $\lambda \to 0$. \figref{fig:virtual_loops} illustrates typical Feynman diagrams contributing to the amplitude for a generic QED hard scattering process. 
In the second diagram, there are several soft virtual photon insertions, joining the external lines associated with the incoming and outgoing hard charged particles. When such insertions are present, the Feynman diagram is infrared logarithmically divergent.

The vanishing of the generic hard  amplitudes due to the exponentiation of the virtual infrared divergences, can also be attributed to constraints and conservation laws associated with large gauge transformations \cite{Lysov:2014csa,He:2014cra,Kapec:2015ena,Campiglia:2015qka,Kapec:2017tkm,Strominger:2017zoo,Gabai:2016kuf,Mohd:2014oja}. These large gauge transformations, which approach angle-dependent constants at infinity, act non-trivially on the asymptotic charged particle states \cite{Strominger:2017zoo}. 
As a consequence, S-matrix elements between such hard states vanish, since the corresponding scattering processes fail to satisfy the infinite number of conservation laws. 
Non-vanishing S-matrix elements are possible between states carrying trivial soft charge \cite{Gabai:2016kuf}. That is, states that are invariant under large gauge transformations. Such states require the dressing of the incoming and outgoing charged  particles with clouds containing an infinite number of soft photons.  

Real soft divergences, on the other hand, appear in radiation-emitting processes. For a scattering process accompanied by the emission of a single photon of energy $\omega_k$, the amplitude can be systematically expanded, in the low-energy limit $\omega_k\to0$, in powers of $\omega_k$. The leading singular term in this expansion is governed by the (leading) soft-photon theorem. Concretely, given a scattering process $\alpha\rightarrow\beta$, the S-matrix amplitude describing the emission of a soft photon of momentum $k$ and polarization $\epsilon_r^*(\vec{k})$ is given by \cite{Bloch:1937pw,Low:1954kd,Low:1958sn, Burnett:1967km}
\begin{equation}\label{eq:leading_soft_thm}
    S_{\beta\gamma,\alpha}=
    \bigg[
        \sum_{i\in\beta}
            e_i\frac{p_i\cdot\epsilon^*_r(\vec{k})}
                {p_i\cdot k}
        -
        \sum_{i\in\alpha}
            e_i\frac{p_i\cdot\epsilon^*_r(\vec{k})}
                {p_i\cdot k}
    \bigg]\
    S_{\beta\alpha}
    +...
\end{equation}
where $S_{\beta\alpha}$ denotes the amplitude of the hard scattering process. Here, $\alpha=\{e_i,\Vec{p}_i,s_i\}$ collectively denotes the charges, momenta, and spin labels of the incoming particles, and similarly for the collective index $\beta$ regarding the outgoing particles. As seen in \eqref{eq:leading_soft_thm}, in the soft limit the single photon-emitting amplitude factorizes. 
The universal soft factor diverges as $1/\omega_k$ in the soft limit 
and depends on the external hard particles only through their charges $e_i$ and momenta $p_i$. Viewed from the symmetry perspective, the leading soft-photon theorem can be realized as a Ward identity associated with large 
gauge transformations \cite{He:2014cra,Campiglia:2015qka,Kapec:2015ena,Mohd:2014oja}. The ellipses stand for subleading terms in the soft-momentum expansion, whose exact form depends on 
the order in perturbation theory at which $S_{\beta\gamma,\alpha}$ is evaluated (see also \figref{fig:rad_emitting_dependence_table}).

At tree level, the subleading term in the soft expansion, first determined by Low \cite{Low:1958sn} and Burnett and Kroll  \cite{Burnett:1967km}, remains constant as $\omega_k$ goes to zero. This subleading contribution to the single photon-emitting amplitude plays an important role in the derivation of the subleading soft dressings for the asymptotic charged states. 
Its precise form can be deduced from the ${\mathcal{O}}(\omega_k^0)$ terms in the expression for the soft factor below:
\begin{equation}\label{eq:subleading_soft_thm}
    S_{\beta\gamma,\alpha}^{(\text{tree})}=
    \bigg[
        \sum_{i\in\beta}
            e_i\frac{\epsilon^*_{\lambda\mu}(\vec{k})}
            {p_i\cdot k}\
            \big(
                p^{\mu}_i+
                ik_{\nu}\bar{J}^{\mu\nu}_i
            \big)
        -
        \sum_{i\in\alpha}
            e_i\frac{\epsilon^*_{\lambda\mu}(\vec{k})}
            {p_i\cdot k}\
            \big(
                p^{\mu}_i+
                ik_{\nu}J^{\mu\nu}_i
            \big)
    \bigg]\
    S_{\beta\alpha}^{\text{(tree)}}+
    \mathcal{O}(\omega_k)
\end{equation} 
At next-to-leading order in the soft-momentum expansion, the tree-level amplitude continues to factorize, with the soft factor including the leading and subleading contributions. The subleading contributions also depend on 
the total angular momentum of the incoming and outgoing charged particles, as indicated by the appearance of the corresponding operators $\bar{J}_i^{\mu\nu}$ (for the outgoing particles) and $J_i^{\mu\nu}$ (for incoming particles)\footnote{
    In presenting the subleading soft-photon theorem in \eqref{eq:subleading_soft_thm}, we assume that no anti-particles are present in the scattering process described by $S_{\beta\gamma,\alpha}^{(\text{tree})}$, neither incoming nor outgoing. For 
    cases involving anti-particles, see \secref{sec:explicit} (or see also Appendix B in \cite{Christodoulou:2025jus}).
}.
See \secref{sec:explicit} for the explicit expressions of the total angular momentum operators.

At loop level, the subleading term in the soft expansion also factorizes \cite{Sahoo:2018lxl}. However, its exact form is beyond the scope of this work, since our aim is to construct the subleading dressings to first order in the QED coupling constant. We note, however, that the subleading contribution to the loop-level scattering amplitude diverges logarithmically in the soft limit:
\begin{equation}
    S_{\beta\gamma,\alpha}^{(\text{loop})}=
    \bigg[
        \sum_{i\in\beta}
            e_i\frac{p_i\cdot\epsilon^*_r(\vec{k})}
                {p_i\cdot k}
        -
        \sum_{i\in\alpha}
            e_i\frac{p_i\cdot\epsilon^*_r(\vec{k})}
                {p_i\cdot k}
    \bigg]\
    S_{\beta\alpha}^{\text{(loop)}}
    + \mathcal{O}(\ln\omega_k)
\end{equation}
See \cite{He:2014bga,Mao:2017wvx,Bern:2014oka,Laddha:2018myi,Saha:2019tub,Sahoo:2019yod,Sahoo:2020ryf,Delisle:2020uui,Krishna:2023fxg} for extensive treatments.

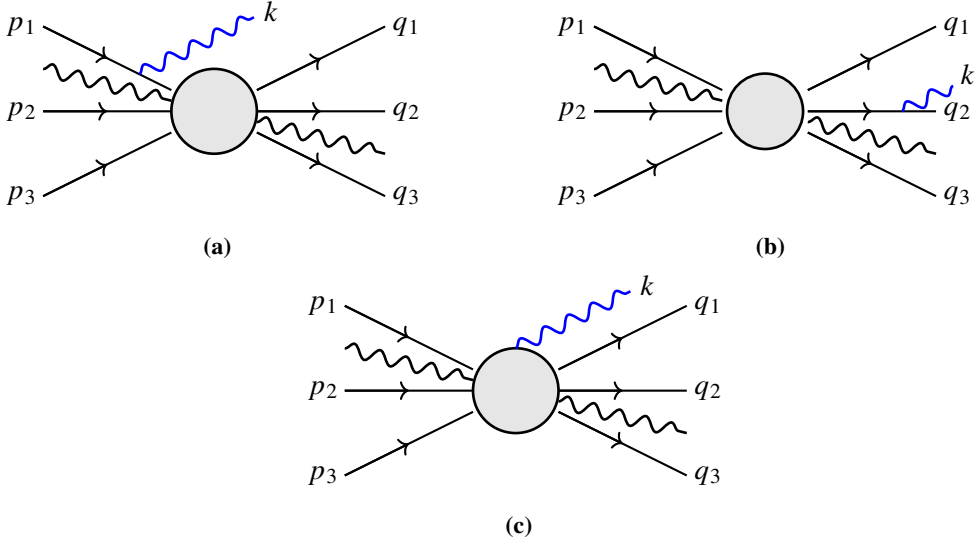
\begin{figure}[!t]
    \begin{subfigure}[b]{0.475\textwidth}
    \centering
        \begin{tikzpicture}
[line width=1pt, 
  fermion/.style={thick},  
  boson/.style={decorate, decoration={snake}, draw=black},
  softboson/.style={decorate, decoration={snake}, draw=blue},
  halffermion/.style={thick, ->}, scale=1.125]
    
  \filldraw[fill=gray!20, draw=black] (0,0) circle (0.5);

  \draw[fermion] (-2,1) -- (-0.5,0.25);
    \draw[halffermion] (-2,1) -- (-1.25,0.625);
    \node at (-2.25,1) {$p_1$};
  \draw[fermion] (-2,0) -- (-0.5,0.0);
    \draw[halffermion] (-2,0) -- (-1.25,0);
    \node at (-2.25,0) {$p_2$};
  \draw[fermion] (-2,-1) -- (-0.5,-0.25);
    \draw[halffermion] (-2,-1) -- (-1.25,-0.625);
    \node at (-2.25,-1) {$p_3$};

  \draw[fermion] (0.5,0.25) -- (2,1);
    \draw[halffermion] (0.5,0.25) -- (1.25,0.625);
    \node at (2.25,1) {$q_1$};
  \draw[fermion] (0.5,0.0) -- (2,0);
    \draw[halffermion] (0.5,0.0) -- (1.25,0);
    \node at (2.25,0) {$q_2$};
  \draw[fermion] (0.5,-0.25) -- (2,-1);
    \draw[halffermion] (0.5,-0.25) -- (1.25,-0.625);
    \node at (2.25,-1) {$q_3$};

  \draw[boson] (0.5,-0.125) -- (2,-0.5);
  \draw[boson] (-2,0.5) -- (-0.5,0.125);

  \draw[softboson] (-0.875, 0.4375) -- (0.466, 1.108);
    \node at (0.666, 1.182) {$k$};

\end{tikzpicture}
    \caption{}
    \end{subfigure}
    \begin{subfigure}[b]{0.475\textwidth}
    \centering
        \begin{tikzpicture}
[line width=1pt, 
  fermion/.style={thick},  
  boson/.style={decorate, decoration={snake}, draw=black},
  softboson/.style={decorate, decoration={snake}, draw=blue},
  halffermion/.style={thick, ->}, scale=1.125]

  \node[draw=black, fill=gray!20, circle, minimum size=1cm, inner sep=0pt] (blob) at (0,0) {};

  \draw[fermion] (-2,1) -- (-0.5,0.25);
    \draw[halffermion] (-2,1) -- (-1.25,0.625);
    \node at (-2.25,1) {$p_1$};
  \draw[fermion] (-2,0) -- (-0.5,0.0);
    \draw[halffermion] (-2,0) -- (-1.25,0);
    \node at (-2.25,0) {$p_2$};
  \draw[fermion] (-2,-1) -- (-0.5,-0.25);
    \draw[halffermion] (-2,-1) -- (-1.25,-0.625);
    \node at (-2.25,-1) {$p_3$};

  \draw[fermion] (0.5,0.25) -- (2,1);
    \draw[halffermion] (0.5,0.25) -- (1.25,0.625);
    \node at (2.25,1) {$q_1$};
  \draw[fermion] (0.5,0.0) -- (2,0);
    \draw[halffermion] (0.5,0.0) -- (1.25,0);
    \node at (2.25,0) {$q_2$};
  \draw[fermion] (0.5,-0.25) -- (2,-1);
    \draw[halffermion] (0.5,-0.25) -- (1.25,-0.625);
    \node at (2.25,-1) {$q_3$};

  \draw[boson] (0.5,-0.125) -- (2,-0.5);
  \draw[boson] (-2,0.5) -- (-0.5,0.125);

  \draw[softboson] (1.6, 0) -- (2.2, 0.3);
    \node at (2.3705, 0.43525) {$k$};

\end{tikzpicture}
    \caption{}
    \end{subfigure}
    \\
    \begin{subfigure}[b]{\textwidth}
    \centering
    \begin{subfigure}[b]{0.475\textwidth}
    \centering
        \begin{tikzpicture}
[line width=1pt, 
  fermion/.style={thick},  
  boson/.style={decorate, decoration={snake}, draw=black},
  softboson/.style={decorate, decoration={snake}, draw=blue},
  halffermion/.style={thick, ->}, scale=1.125]
    
  \filldraw[fill=gray!20, draw=black] (0,0) circle (0.5);

  \draw[fermion] (-2,1) -- (-0.5,0.25);
    \draw[halffermion] (-2,1) -- (-1.25,0.625);
    \node at (-2.25,1) {$p_1$};
  \draw[fermion] (-2,0) -- (-0.5,0.0);
    \draw[halffermion] (-2,0) -- (-1.25,0);
    \node at (-2.25,0) {$p_2$};
  \draw[fermion] (-2,-1) -- (-0.5,-0.25);
    \draw[halffermion] (-2,-1) -- (-1.25,-0.625);
    \node at (-2.25,-1) {$p_3$};

  \draw[fermion] (0.5,0.25) -- (2,1);
    \draw[halffermion] (0.5,0.25) -- (1.25,0.625);
    \node at (2.25,1) {$q_1$};
  \draw[fermion] (0.5,0.0) -- (2,0);
    \draw[halffermion] (0.5,0.0) -- (1.25,0);
    \node at (2.25,0) {$q_2$};
  \draw[fermion] (0.5,-0.25) -- (2,-1);
    \draw[halffermion] (0.5,-0.25) -- (1.25,-0.625);
    \node at (2.25,-1) {$q_3$};

  \draw[boson] (0.5,-0.125) -- (2,-0.5);
  \draw[boson] (-2,0.5) -- (-0.5,0.125);

  \draw[softboson] (0, 0.5) -- (1.341, 1.1705);
    \node at (1.541, 1.2405) {$k$};

\end{tikzpicture}
    \caption{}
    \end{subfigure}
    \end{subfigure}
    \caption{
        Diagrams depicting single soft-photon emission during a generic QED scattering process. The interactions are collectively represented by the blob. Diagrams (a) and (b) describe soft-photon emission from outgoing and incoming charged particles, respectively, while diagram (c) corresponds to soft-photon emission from a virtual charged particle.
        }
    \label{fig:photon_emission}
\end{figure}

The leading contribution to the soft-photon theorem is common to all orders in perturbation theory. The subleading contribution, on the other hand, is not: at tree level, it is $\mathcal{O}(\omega_k^0)$, while at loop level it is $\mathcal{O}(\ln\omega_k)$. Hence, as higher orders in perturbation theory are taken into account, additional infrared divergences in soft-photon-emitting amplitudes appear. \figref{fig:photon_emission} illustrates all possible ways in which an additional soft photon may be emitted in a generic scattering process. All these ways must be taken into account in order to derive the leading and subleading soft-photon theorems at tree level.

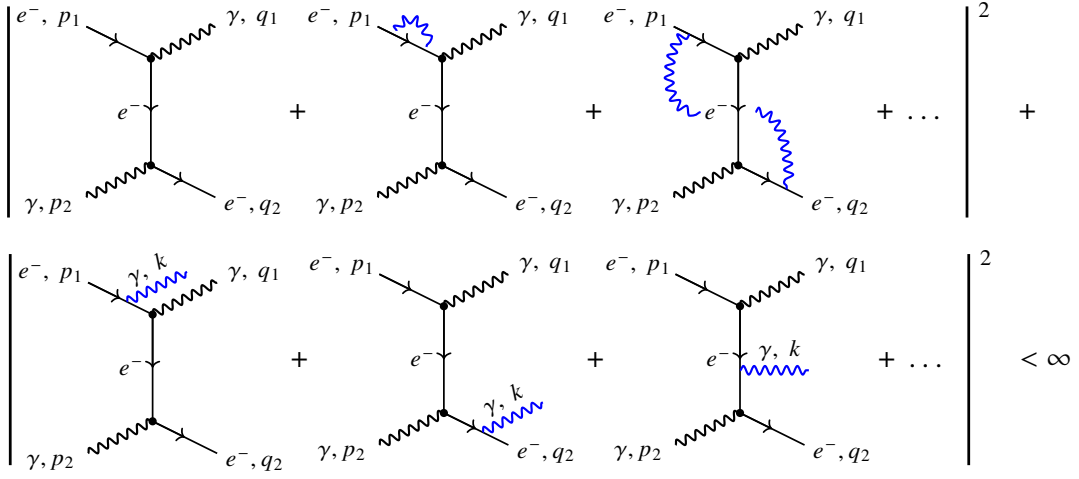
\begin{figure}[!t]
    \centering
    \begin{minipage}[c]{\linewidth}
    \vspace{0.5em}
    \centering
        \begin{minipage}[c]{0.05\linewidth}
            \centering
            \resizebox{\linewidth}{!}{\begin{tikzpicture}[line width=1pt,
  fermion/.style={thick},
  boson/.style={decorate, decoration={snake, amplitude=2pt, segment length=5pt}, draw=black},
  halffermion/.style={thick, ->},scale=0.7]

  \path[use as bounding box] (-0.2,-1.6) rectangle (0.5,1.6);

  \draw[fermion] (0,-1.55) -- (0,1.55);
  \node at (0.25,1.5) {\phantom{\tiny{$2$}}};

\end{tikzpicture}}
        \end{minipage}
        \hspace{-2em}
        \begin{minipage}[c]{0.25\linewidth}
            \centering
            \resizebox{\linewidth}{!}{\begin{tikzpicture}[line width=1pt, 
  fermion/.style={thick},  
  boson/.style={decorate, decoration={snake, amplitude=2pt, segment length=5pt}, draw=black},
  halffermion/.style={thick, ->},scale=0.7]

  \draw[fermion] (0.5,2) -- (2,1.25);    
  \draw[boson] (0.5,-2) -- (2,-1.25);  
  \draw[boson] (2,1.25) -- (3.5,2);    
  \draw[fermion] (2,-1.25) -- (3.5,-2);  

  \draw[fermion] (2,1.25) -- (2,-1.25) node[midway, left] {$e^-$};
  \draw[halffermion] (2,1.25) -- (2,0); 
  
  \filldraw[black] (2,1.25) circle (2pt);
  \filldraw[black] (2,-1.25) circle (2pt);

  \draw[halffermion] (0.5,2) -- (1.25,1.625);    
  \draw[halffermion] (2,-1.25) -- (2.75,-1.625);  

  \node at (-0.3,2.2) {$e^-,\ p_1\ $};
  \node at (-0.3,-2.2) {$\gamma, p_2\ $};
  \node at (4.3,2.2) {$\ \gamma,\ q_1$};
  \node at (4.3,-2.2) {$\ e^-, q_2$};

  
\end{tikzpicture}}
        \end{minipage}
        $\!\!\!+\!\!\!$
        \begin{minipage}[c]{0.25\linewidth}
            \centering
            \resizebox{\linewidth}{!}{\begin{tikzpicture}[line width=1pt, 
  fermion/.style={thick},  
  boson/.style={decorate, decoration={snake, amplitude=2pt, segment length=5pt}, draw=black},
  halffermion/.style={thick, ->},scale=0.7]

  \draw[fermion] (0.5,2) -- (2,1.25);    
  \draw[boson] (0.5,-2) -- (2,-1.25);  
  \draw[boson] (2,1.25) -- (3.5,2);    
  \draw[fermion] (2,-1.25) -- (3.5,-2);  

  \draw[fermion] (2,1.25) -- (2,-1.25) node[midway, left] {$e^-$};
  \draw[halffermion] (2,1.25) -- (2,0); 
  
  \filldraw[black] (2,1.25) circle (2pt);
  \filldraw[black] (2,-1.25) circle (2pt);

  \draw[halffermion] (0.5,2) -- (1.25,1.625);    
  \draw[halffermion] (2,-1.25) -- (2.75,-1.625);  

  \node at (-0.3,2.2) {$e^-,\ p_1\ $};
  \node at (-0.3,-2.2) {$\gamma, p_2\ $};
  \node at (4.3,2.2) {$\ \gamma,\ q_1$};
  \node at (4.3,-2.2) {$\ e^-, q_2$};

  \draw[boson, draw=blue, shift={(0.425,-0.295)}, rotate=-10] 
  (0.125,2.25) 
  .. controls (0.25,2.75) and (0.875,2.5) 
  .. (0.875,2);
  
\end{tikzpicture}}
        \end{minipage}
        $\!\!+\!\!\!$
        \begin{minipage}[c]{0.25\linewidth}
            \centering
            \resizebox{\linewidth}{!}{\begin{tikzpicture}[line width=1pt, 
  fermion/.style={thick},  
  boson/.style={decorate, decoration={snake, amplitude=2pt, segment length=5pt}, draw=black},
  halffermion/.style={thick, ->},scale=0.7]

  \draw[fermion] (0.5,2) -- (2,1.25);    
  \draw[boson] (0.5,-2) -- (2,-1.25);  
  \draw[boson] (2,1.25) -- (3.5,2);    
  \draw[fermion] (2,-1.25) -- (3.5,-2);  

  \draw[fermion] (2,1.25) -- (2,-1.25) node[midway, left] {$e^-$};
  \draw[halffermion] (2,1.25) -- (2,0); 
  
  \filldraw[black] (2,1.25) circle (2pt);
  \filldraw[black] (2,-1.25) circle (2pt);

  \draw[halffermion] (0.5,2) -- (1.25,1.625);    
  \draw[halffermion] (2,-1.25) -- (2.75,-1.625);  

  \node at (-0.3,2.2) {$e^-,\ p_1\ $};
  \node at (-0.3,-2.2) {$\gamma, p_2\ $};
  \node at (4.3,2.2) {$\ \gamma,\ q_1$};
  \node at (4.3,-2.2) {$\ e^-, q_2$};

  \draw[boson, draw=blue] (0.875,1.8125) 
  .. controls (0.5,1.8125) and (0.125,0) 
  .. (1.125,0);
  \draw[boson, draw=blue] (2.4,0) 
  .. controls (3,0) and (3.125,-1.125)
  .. (3.125,-1.8125);
  
\end{tikzpicture}}
        \end{minipage}
        $\!\!\!+~\dots$
        \begin{minipage}[c]{0.05\linewidth}
            \centering
            \resizebox{\linewidth}{!}{\begin{tikzpicture}[line width=1pt,
  fermion/.style={thick},
  boson/.style={decorate, decoration={snake, amplitude=2pt, segment length=5pt}, draw=black},
  halffermion/.style={thick, ->},scale=0.7]

  \path[use as bounding box] (-0.2,-1.6) rectangle (0.5,1.6);

  \draw[fermion] (0,-1.55) -- (0,1.55);
  \node at (0.25,1.5) {{\tiny{$2$}}};

\end{tikzpicture}}
        \end{minipage}
        $+$
        \hspace{1em}
    \end{minipage}
    \\
    \begin{minipage}[c]{\linewidth}
    \vspace{0.5em}
    \centering
        \begin{minipage}[c]{0.05\linewidth}
            \centering
            \resizebox{\linewidth}{!}{\begin{tikzpicture}[line width=1pt,
  fermion/.style={thick},
  boson/.style={decorate, decoration={snake, amplitude=2pt, segment length=5pt}, draw=black},
  halffermion/.style={thick, ->},scale=0.7]

  \path[use as bounding box] (-0.2,-1.6) rectangle (0.5,1.6);

  \draw[fermion] (0,-1.55) -- (0,1.55);
  \node at (0.25,1.5) {\phantom{\tiny{$2$}}};

\end{tikzpicture}}
        \end{minipage}
        \hspace{-2em}
        \begin{minipage}[c]{0.25\linewidth}
            \centering
            \resizebox{\linewidth}{!}{\begin{tikzpicture}[line width=1pt, 
  fermion/.style={thick},  
  boson/.style={decorate, decoration={snake, amplitude=2pt, segment length=5pt}, draw=black},
  halffermion/.style={thick, ->},scale=0.7]

  \draw[fermion] (0.5,2) -- (2,1.25);    
  \draw[boson] (0.5,-2) -- (2,-1.25);  
  \draw[boson] (2,1.25) -- (3.5,2);    
  \draw[fermion] (2,-1.25) -- (3.5,-2);  

  \draw[fermion] (2,1.25) -- (2,-1.25) node[midway, left] {$e^-$};
  \draw[halffermion] (2,1.25) -- (2,0); 
  
  \filldraw[black] (2,1.25) circle (2pt);
  \filldraw[black] (2,-1.25) circle (2pt);

  \draw[halffermion] (0.5,2) -- (1.25,1.625);    
  \draw[halffermion] (2,-1.25) -- (2.75,-1.625);  

  \node at (-0.3,2.2) {$e^-,\ p_1\ $};
  \node at (-0.3,-2.2) {$\gamma, p_2\ $};
  \node at (4.3,2.2) {$\ \gamma,\ q_1$};
  \node at (4.3,-2.2) {$\ e^-, q_2$};


  \coordinate (emit) at ($(0.5,2)!0.6!(2,1.25)$);
  \draw[boson, draw=blue]
    (emit) -- ++(1.4,0.7)     
    node[midway, above, sloped] {$\gamma,\ k$};
  
\end{tikzpicture}}
        \end{minipage}
        $\!\!\!+\!\!\!$
        \begin{minipage}[c]{0.25\linewidth}
            \centering
            \resizebox{\linewidth}{!}{\begin{tikzpicture}[line width=1pt, 
  fermion/.style={thick},  
  boson/.style={decorate, decoration={snake, amplitude=2pt, segment length=5pt}, draw=black},
  halffermion/.style={thick, ->},scale=0.7]

  \draw[fermion] (0.5,2) -- (2,1.25);    
  \draw[boson] (0.5,-2) -- (2,-1.25);  
  \draw[boson] (2,1.25) -- (3.5,2);    
  \draw[fermion] (2,-1.25) -- (3.5,-2);  

  \draw[fermion] (2,1.25) -- (2,-1.25) node[midway, left] {$e^-$};
  \draw[halffermion] (2,1.25) -- (2,0); 
  
  \filldraw[black] (2,1.25) circle (2pt);
  \filldraw[black] (2,-1.25) circle (2pt);

  \draw[halffermion] (0.5,2) -- (1.25,1.625);    
  \draw[halffermion] (2,-1.25) -- (2.75,-1.625);  

  \node at (-0.3,2.2) {$e^-,\ p_1\ $};
  \node at (-0.3,-2.2) {$\gamma, p_2\ $};
  \node at (4.3,2.2) {$\ \gamma,\ q_1$};
  \node at (4.3,-2.2) {$\ e^-, q_2$};


  \coordinate (emit2) at ($(2,-1.25)!0.6!(3.5,-2)$);
  \draw[boson, draw=blue]
    (emit2) -- ++(1.4,0.7)
    node[midway, above, sloped] {$\gamma,\ k$};
  
\end{tikzpicture}}
        \end{minipage}
        $\!\!+\!\!\!$
        \begin{minipage}[c]{0.25\linewidth}
            \centering
            \resizebox{\linewidth}{!}{\begin{tikzpicture}[line width=1pt, 
  fermion/.style={thick},  
  boson/.style={decorate, decoration={snake, amplitude=2pt, segment length=5pt}, draw=black},
  halffermion/.style={thick, ->},scale=0.7]

  \draw[fermion] (0.5,2) -- (2,1.25);    
  \draw[boson] (0.5,-2) -- (2,-1.25);  
  \draw[boson] (2,1.25) -- (3.5,2);    
  \draw[fermion] (2,-1.25) -- (3.5,-2);  

  \draw[fermion] (2,1.25) -- (2,-1.25) node[midway, left] {$e^-$};
  \draw[halffermion] (2,1.25) -- (2,0); 
  
  \filldraw[black] (2,1.25) circle (2pt);
  \filldraw[black] (2,-1.25) circle (2pt);

  \draw[halffermion] (0.5,2) -- (1.25,1.625);    
  \draw[halffermion] (2,-1.25) -- (2.75,-1.625);  

  \node at (-0.3,2.2) {$e^-,\ p_1\ $};
  \node at (-0.3,-2.2) {$\gamma, p_2\ $};
  \node at (4.3,2.2) {$\ \gamma,\ q_1$};
  \node at (4.3,-2.2) {$\ e^-, q_2$};


  \coordinate (emit3) at ($(2,1.25)!0.6!(2,-1.25)$);
  \draw[boson, draw=blue]
    (emit3) -- ++(1.6,0)
    node[midway, above] {$\ \ \gamma,\ k$};
  
\end{tikzpicture}}
        \end{minipage}
        $\!\!\!+~\dots$
        \begin{minipage}[c]{0.05\linewidth}
            \centering
            \resizebox{\linewidth}{!}{\begin{tikzpicture}[line width=1pt,
  fermion/.style={thick},
  boson/.style={decorate, decoration={snake, amplitude=2pt, segment length=5pt}, draw=black},
  halffermion/.style={thick, ->},scale=0.7]

  \path[use as bounding box] (-0.2,-1.6) rectangle (0.5,1.6);

  \draw[fermion] (0,-1.55) -- (0,1.55);
  \node at (0.25,1.5) {{\tiny{$2$}}};

\end{tikzpicture}}
        \end{minipage}
        $<\infty$
    \end{minipage}
    \caption{
        A schematic representation of the cancellation of virtual infrared divergences against real soft infrared divergences at the level of inclusive rates. Representative examples of Feynman diagrams depicting Compton scattering are used. 
    }
    \label{fig:BN_cancellation}
\end{figure}

The appearance of infrared divergences
does not obstruct collider predictions. As suggested by Bloch and Nordsieck \cite{Bloch:1937pw}, inclusive cross sections or rates remain infrared-finite and nonvanishing order by order in perturbation theory, even though S-matrix elements between Fock-basis states with a finite number of soft photons are infrared divergent. See \figref{fig:BN_cancellation} for a pictorial representation of the Bloch-Nordsieck cancellation in Compton scattering at sixth order 
in the coupling constant. Such inclusive treatments, however, fail to describe the correct asymptotic particle states 
and do not provide a well-defined S-matrix for studying scattering in QED.

\section{The dressed state formalism}
    \label{sec:dressed_states}
    Even though infrared divergences do not prevent us from making explicit collider predictions, one 
would still expect to have a well-defined, infrared-finite 
S-matrix
in gauge theories and gravity. 
Faddeev and Kulish \cite{Kulish:1970ut} addressed this question by utilizing the 
correct asymptotic states, in which charged incoming and outgoing particles are accompanied by coherent clouds of soft photons. These asymptotic states, referred to as dressed states, are invariant under large gauge transformations.

Faddeev and Kulish used the fact that 
in a scattering process, well before and well after the scattering takes place, the QED interactions are not completely turned off. Indeed, it can be shown that the slowly decaying parts of the QED potential in the interaction picture, which persist in the limit $|t|\to\infty$, affect the asymptotic dynamics. This nonvanishing contribution to the QED potential is given by: 
\begin{equation}\label{eq:Vas}
    V_{as}^I(t)=-e\!
        \int\frac{d^3p}{(2\pi)^3(2\omega_p)}\ 
        \rho(\Vec{p})\ p^{\mu}
        \int\frac{d^3k}{(2\pi)^3(2\omega_k)}\ 
        \big[
            \alpha_{\mu}(\Vec{k})
                e^{ip\cdot kt/p^0}+
            \text{h.c.}
        \big]
\end{equation}
where $\rho(\Vec{p})$ is the charge density operator
\begin{equation}
    \rho(\Vec{p})=
        \sum_s
        \big[
            b^{\dagger}_s(\Vec{p})b_s(\Vec{p})- 
            d^{\dagger}_s(\Vec{p})d_s(\Vec{p})
        \big]
\end{equation}
with $b^{\dagger}_s(\Vec{p})$ and $d^{\dagger}_s(\Vec{p})$ creating electrons and positrons\footnote{In case more charged fermions are included, we must include their contributions to the charge density operator, as well.} with momentum $\vec{p}$, spin polarization $s$, and energy $\omega_p=\sqrt{|\vec{p}|^2+m^2}$. $a_r(\vec{k})$ annihilates a photon with momentum $\vec{k}$, polarization vector $\epsilon_r^{\mu}(\vec{k}),\ r=0,\ 1,\ 2,\ 3$, and energy $\omega_k=|\vec{k}|$, and
\begin{equation}
    a^{\mu}(\vec{k})=
        \sum_{\lambda}\epsilon_{r}^{\mu}(\vec{k})~
        a_{r}(\vec{k})
\end{equation}
Due to the highly oscillatory nature of the integrand in \eqref{eq:Vas} in the limit $|t|\to\infty$, the integral is dominated by and remains non-vanishing due to the very low-energy, soft-photon contributions.

\subsection{Faddeev-Kulish dressed states}
    Since the asymptotic dynamics of QED differs from that of the free theory,
    it is not justified to define S-matrix elements in terms of the eigenstates of the free Hamiltonian. Instead, S-matrix elements should be defined in terms of eigenstates of the asymptotic Hamiltonian that captures the long-range dynamics of the QED interaction potential at large asymptotic times. As shown by Fadeev and Kulish, these eigenstates can be obtained by acting on the Fock-basis states with a unitary operator, called the dressing operator. Starting with a multiparticle Fock-basis state $\ket{\alpha}$, where $\alpha=\{e_i,\Vec{p}_i,s_i\}$, the Fadeev-Kulish dressed state is given by  \cite{Kulish:1970ut,Tomaras_2020,Gabai:2016kuf,Gomez:2017rau}
\begin{equation}\label{eq:dressed_state}
        \ket{\alpha}_d=
            e^{R_f}\ket{\alpha}
    \end{equation}
  where the anti-hermitian exponent of the 
  dressing operator $e^{R_f}$ is given by
    \begin{equation}\label{eq:R_f}
        R_f=
            \int\frac{d^3p}{(2\pi)^3(2\omega_p)}\ 
                \rho(\Vec{p})
            \int
                \reallywidetilde{d^3k}\
                \big[
                    f(\Vec{p},\Vec{k})
                        \cdot
                    \alpha^{\dagger}(\Vec{k})-
                    f^*(\Vec{p},\Vec{k})
                        \cdot
                    \alpha(\Vec{k})
                \big]
    \end{equation}
    The dot product between the dressing function $f^{\mu}(\vec{p},\vec{k})$ and the (soft) photon creation operator is defined as
    \begin{equation}\label{eq:dot_prod_leading}
        f(\Vec{p},\Vec{k})
            \cdot
        \alpha^{\dagger}(\Vec{k})=
        \sum_{r}
            f^{\mu}(\Vec{p},\Vec{k})~
            \epsilon^*_{r\mu}(\Vec{k})~
            \alpha_{\lambda}^{\dagger}(\Vec{k})
    \end{equation}
    with
    \begin{equation}
        f^{\mu}(\Vec{p},\Vec{k})=
            \Big(
                \frac{p^{\mu}}{p\cdot k}-
                c^{\mu}
            \Big)e^{-ip\cdot kt/p^0}
        ,\hspace{1em}
        c^{\mu}=\frac{1}{2k^0}(-1,\hat{k})
    \end{equation}
    The 
    phase space integral measure is given by
    \begin{equation}
        \int\reallywidetilde{d^3k}=
        \frac{1}{2}
        \int
            \frac{d\Omega_k}
                {(2\pi)^3}
        \int_{\lambda}^{E_d}
            \omega_kd\omega_k
    \end{equation}
    and includes integration over the direction of the unit vector $\hat{k}$ and over the magnitude of $\vec{k}$, which is equal to the photon energy, $\omega_k=|\vec{k}|$. The integration limits for $\omega_k$ are fixed by the infrared regulator $\lambda$, which will be taken to zero, and the infrared energy scale $E_d$, which characterizes the mean energy of the soft photons in the clouds that accompany the charged particles.  
    We fix $E_d$ to be sufficiently small\footnote{
        We take $E_d<\Lambda$, where $\Lambda$ is the infrared energy scale characterizing virtual soft photons.
    } in order to match the low-energy contributions to the asymptotic potential \cite{Gomez:2018war}. 

    The dressing function contains the leading pole that appears also in the soft-photon theorem. As we will see, this ensures the infrared-finiteness of dressed elastic amplitudes, as well as the removal of the leading pole in dressed soft-photon-emitting amplitudes. 
    The null vector $c^{\mu}$ is introduced so that the dressing function is transverse. The dressing operator in \eqref{eq:R_f} creates cloud photons with energy between the infrared regulator $\lambda$ and the characteristic energy scale $E_d$.

\begin{figure}
        \centering
        \begin{minipage}[c]{\linewidth}
        \vspace{0.5em}
        \centering
            \begin{minipage}[c]{0.425\linewidth}
                \centering
                \resizebox{0.75\linewidth}{!}{\begin{tikzpicture}
[line width=1pt, 
  fermion/.style={thick},  
  boson/.style={decorate, decoration={snake}, draw=black},
  softboson/.style={decorate, decoration={snake}, draw=blue},
  halffermion/.style={thick, ->}, 
  scale=1.5]

    \filldraw[fill=gray!20, draw=black] (-2,0) circle (0.5);

    \draw[fermion] (-2.25,0.4330) -- (-2.884,1.530);
        \draw[halffermion] (-2.884,1.530) -- (-2.567,0.9815);
        \node at (-3.45, 1.60) {$p_1$};   
    \draw[fermion] (-2.3536,0.3536) -- (-3.25,1.25);
        \draw[halffermion] (-3.25,1.25) -- (-2.8018,0.8018);
        \node at (-3.8, 1.325) {$p_2$};    
    \draw[fermion] (-2.4330,0.25) -- (-3.530,0.884);
        \draw[halffermion] (-3.530,0.884) -- (-2.9815,0.567);
        \node at (-4.05, 0.884) {$p_3$};  
    \draw[fermion] (-2.4330,-0.25) -- (-3.530,-0.884);
        \draw[halffermion] (-3.530,-0.884) -- (-2.9815,-0.567);
        \node at (-4.1, -0.884) {$p_{n-2}$}; 
    \draw[fermion] (-2.3536,-0.3536) -- (-3.25,-1.25);
        \draw[halffermion] (-3.25,-1.25) -- (-2.8018,-0.8018);
        \node at (-3.85, -1.3) {$p_{n-1}$};   
    \draw[fermion] (-2.25,-0.4330) -- (-2.884,-1.530);
        \draw[halffermion] (-2.884,-1.530) -- (-2.567,-0.9815);
        \node at (-3.45, -1.625) {$p_n$};     

    \fill (-2.9, 0.225) circle (1pt);
    \fill (-3, 0) circle (1pt);
    \fill (-2.9, -0.225) circle (1pt);    

    \draw[boson] (-1.75, 0.4330) -- (-1.5387,0.7987);
    \draw[boson] (-1.6464, 0.3536) -- (-1.3476,0.6197);
    \draw[fermion] (-1.567,0.25) -- (-1.2013,0.4613);
    \draw[halffermion] (-1.567,0.25) -- (-1.384,0.3557);
    \draw[boson] (-1.567, -0.25) -- (-1.2013,-0.4613);
    \draw[boson] (-1.6464, -0.3536) -- (-1.3476,-0.6197);
    \draw[fermion] (-1.75, -0.4330) -- (-1.5387,-0.7987);
        \draw[halffermion] (-1.75, -0.4330) -- (-1.6444,-0.6159);
    \draw[fermion] (-1.5,0) -- (-1.12,0);
        \draw[halffermion] (-1.5,0) -- (-1.31,0);

\end{tikzpicture}}
            \end{minipage}
            $\Rightarrow$
            \begin{minipage}[c]{0.425\linewidth}
                \centering
                \resizebox{0.75\linewidth}{!}{\input{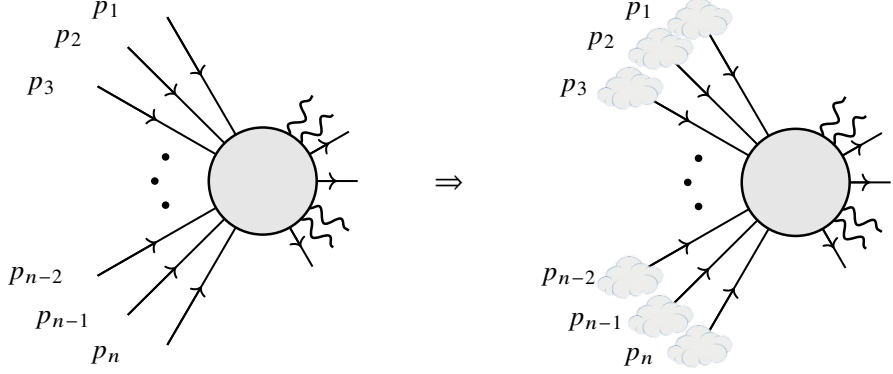}}
            \end{minipage}
        \end{minipage}
        \caption{
            Pictorial representation of the replacement of $n$ incoming charged particle state by the corresponding dressed state in a generic QED scattering process. 
            An analogous replacement can be applied for the $m$ outgoing charged particle state, which has not been displayed in full. 
        }
        \label{fig:Fock_replacement}
    \end{figure}

    The dressed state in \eqref{eq:dressed_state} can be written as a product between a Fock-basis state of the charged particles and a coherent state of soft photons \cite{Tomaras_2020}, as follows:
    \begin{equation}
        \ket{\alpha}_d=
            \ket{\alpha}
            \times
            \ket{f_{\alpha}}
    \end{equation}
    where
    \begin{equation*}
        \ket{f_{\alpha}}=
            \mathcal{N}_{\alpha}\
            e^{\int
                \reallywidetilde{d^3k}\
                    f_{\alpha}(\Vec{k})
                        \cdot
                    \alpha^{\dagger}(\Vec{k})
                }
            \ket{0}
    \end{equation*} 
    and
    \begin{equation*}
        \mathcal{N}_{\alpha}=
            e^{
                -\frac{1}{2}
                \int
                \reallywidetilde{d^3k}\
                f_{\alpha}^*(\Vec{k})
                    \cdot
                f_{\alpha}(\Vec{k})
            }
        ,\hspace{1em}
        f^{\mu}_{\alpha}(\Vec{k})=
            \sum_{i\in\alpha}
            e_i
            \Big(
                \frac{p_i^{\mu}}{p_i\cdot k}-
                c^{\mu}
            \Big)e^{-ip_i\cdot kt/p_i^0}
    \end{equation*}
    Diagrammatically, we illustrate the replacement of the Fock states with the corresponding dressed states in \figref{fig:Fock_replacement}.

    \subsection{The Faddeev-Kulish S-matrix}
    
    In this subsection, we summarize some of the properties of scattering amplitudes in the dressed state formalism. For this purpose, 
    we distinguish between dressed elastic amplitudes, involving no additional soft radiation, and dressed amplitudes involving additional soft-photon emissions, apart from the soft photons already present in the clouds that accompany the external charged particles. 
    The dressed elastic amplitude for the scattering process $\alpha\rightarrow\beta$ is given by the following S-matrix element:
    \begin{equation}
        \tilde{S}_{\beta,\alpha}=
            \tensor[_d]{\braket{\beta
                |S|
            \alpha}}{_d}
    \end{equation}
    If an additional soft photon is emitted during the process $\alpha\rightarrow\beta$, we get
    \begin{equation}
        \tilde{S}_{\beta\gamma,\alpha}=
            \tensor[_d]{\braket{\beta\gamma
                |S|
            \alpha}}{_d}
    \end{equation}
    where $\gamma$ denotes the additional final state photon of momentum $k$ and polarization vector $\epsilon^*_r(\vec{k}),\ r=1,\ 2$. The main properties of these dressed amplitudes are: 
    \begin{itemize}
        \item The dressed elastic amplitudes are free of any infrared divergences \cite{Tomaras_2020}
        \begin{equation}\label{eq:virtual_finite}
        \tilde{S}_{\beta\alpha}=
            \bigg(
                \frac{E_d}{\Lambda}
            \bigg)^
                {\mathcal{B}_{\beta\alpha}}\
            e^{i\phi_{\beta\alpha}}\
            S_{\beta\alpha}^{(\Lambda)}
    \end{equation}
    where $\Lambda$ is the infrared-cutoff scale used to distinguish virtual soft photons, and $\mathcal{B}_{\beta\alpha}$ is given by
    \begin{equation}\label{eq:kinematical_factor}
        \mathcal{B}_{\beta\alpha}=
            -\frac{1}{16\pi^2}
            \sum_{i,j}
            \eta_i\eta_j
            e_ie_j
            \upsilon_{ij}^{-1}
            \ln\bigg(
                \frac{1+\upsilon_{ij}}
                    {1-\upsilon_{ij}}
            \bigg)
    \end{equation}
    with $\eta_i=+1(-1)$ for outgoing (incoming) charged particles, and $\upsilon_{ij}$ the magnitude of the relative velocity of the $j$-th particle with respect to the $i$-th particle, given by
    \begin{equation}
        \upsilon_{ij}=
            \bigg[
                1-\frac{m_i^2m_j^2}{(p_ip_j)^2}
            \bigg]^{1/2}
    \end{equation}
    In addition, $\phi_{\beta\alpha}$ and $S_{\beta\alpha}^{(\Lambda)}$ are a real phase and the undressed S-matrix element stripped from the contributions from virtual soft photons (appearing in \eqref{eq:elastic_vanishing}). It follows from \eqref{eq:virtual_finite} that the dressed amplitude for the process $\alpha\rightarrow\beta$
    remains nonzero and free of any infrared divergences in the limit $\lambda\rightarrow0$, provided that the ratio $E_d/\Lambda$ remains fixed \cite{Tomaras_2020}. Thus, the contributions from the virtual soft photons are effectively removed by the dressing clouds. By choosing $E_d=\Lambda$, the dressed elastic amplitude becomes equal
    to the infrared-finite part of the undressed amplitude, 
    which is constructed using Fock states.
    \item The dressed amplitude for the emission of an additional soft photon is finite, given by \cite{Tomaras_2020}
    \begin{equation*}
        \tilde{S}_{\beta\gamma,\alpha}=
            F_{\beta\alpha}
                \big(
                    \vec{k},
                    \epsilon_r^*(\vec{k})
                \big)
    \end{equation*}
    where the function
    $F_{\beta\alpha}
            \big(
                \vec{k},
                \epsilon_r^*(\vec{k})
            \big)$
    of the momentum and the polarization of the emitted soft photon is free of any infrared divergences 
    as $|\Vec{k}|\rightarrow0$, at least at tree level in perturbation theory \cite{Tomaras_2020}. In fact, to leading order in perturbation theory, this function is smooth as $\lambda,|\vec{k}|\rightarrow0$. 
    (See \secref{sec:sum}
    for discussions regarding higher orders.) Extending the dressings to subleading order in the soft-momentum expansion results in the complete suppression of this soft-photon-emitting amplitude \cite{Choi:2019rlz, Christodoulou:2025jus}. 
    \end{itemize}
\section{Subleading soft radiation during scattering of dressed states}
    \label{sec:explicit}
    Having reviewed the properties of both dressed elastic and soft-photon-emitting amplitudes for generic QED scattering processes, we now consider to apply the formalism in three concrete examples: the scattering of an electron by a heavier muon, Compton scattering, and electron-positron annihilation into two photons. Using these concrete examples, the dependence of the soft factors and, hence, that of the dressing functions on the momenta of the hard particles and the total angular momenta can be made explicit. 

We first present the undressed amplitudes describing soft-photon emission during these three processes at tree level. Following \cite{Choi:2019rlz}, we then construct the corresponding dressed states to subleading order in the soft-momentum expansion and to first order in the QED coupling constant. Finally, we demonstrate that the emission of an additional soft photon with energy less than $E_d$ during scattering of dressed states is completely suppressed
at tree level. This suppression would not be possible without extending the dressing functions of \cite{Dollard:1964jmp,Chung:1965zza,Kibble:1968lka,Kibble:1968npb,Kibble:1968oug,Kibble:1968sfb,Kulish:1970ut}. The subleading corrections to the dressing functions are, therefore, essential.

\subsection{Subleading soft radiation with Fock states}
    Here, we study single soft-photon emission during scattering of Fock states. We use the label $\alpha=\{\mu,\gamma,e\}$ to denote the scattering of an electron by a heavier muon (for $\alpha=\mu$),  electron-photon scattering (for $\alpha=\gamma$) and electron-positron annihilation into two photons (for $\alpha=e$), respectively. First, we define the tree-level elastic amplitude for these three processes, as follows
    \begin{equation}
        S_0
            ^{(\alpha)}
            =
            \braket{q_1, q_2| S |p_1, p_2}_{\rm tree}
    \end{equation}
    with $p_1,\ p_2$ denoting the incoming particles' momenta, and $q_1,\ q_2$ the momenta of the outgoing particles. 
    
    We add a photon of momentum $\vec{k}=|\vec{k}|\hat{k}$, energy $\omega_k=|\vec{k}|$, and polarization vector $\epsilon_r^*(\vec{k}),\ r=1,\ 2$ among the final state particles to describe single soft-photon emission during these three scattering processes. In the soft limit $\omega_k\to0$, the amplitudes factorize into a universal soft factor multiplying the elastic amplitudes $S_0^{(\alpha)}$, in agreement with the subleading soft-photon theorem \cite{Gell-Mann:1954wra,Low:1954kd,Low:1958sn,Burnett:1967km}. The tree-level amplitudes for single soft-photon emission in these three cases are as follows:
    \begin{itemize}
        \item for $e^-(p_1)+\mu^-(p_2)\rightarrow e^-(q_1)+\mu^-(q_2)$:
        \begin{equation}\label{eq:muon}
        \begin{split}
            S_{\text{tree}}^{(\mu)}
                (\omega_k,\hat{k})=
                e \bigg\{
                    &\Big[
                        \frac{q_1\cdot
                                \epsilon_r^*(\vec{k})}
                            {q_1\cdot k}+
                        \frac{q_2\cdot
                                \epsilon_r^*(\vec{k})}
                            {q_2\cdot k}-
                        \frac{p_1\cdot
                                \epsilon_r^*(\vec{k})}
                            {p_1\cdot k}-
                        \frac{p_2\cdot
                                \epsilon_r^*(\vec{k})}
                            {p_2\cdot k}
                    \Big]\\&\!\!+
                    \frac{2k\!\cdot\!(p_1-q_1)}
                        {(p_1-q_1)^2}
                    \Big[
                        \frac{q_1\cdot
                                \epsilon_r^*(\vec{k})}
                            {q_1\cdot k}-
                        \frac{q_2\cdot
                                \epsilon_r^*(\vec{k})}
                            {q_2\cdot k}-
                        \frac{p_1\cdot
                                \epsilon_r^*(\vec{k})}
                            {p_1\cdot k}+
                        \frac{p_2\cdot
                                \epsilon_r^*(\vec{k})}
                            {p_2\cdot k}
                    \Big]\!\\&\!\!+
                    i\epsilon^*_{r\mu}(\vec{k})\ 
                    k_{\nu}
                    \Big[
                        \frac{\bar{S}_{q_1}^{\mu\nu}}
                            {q_1\cdot k}+
                        \frac{\bar{S}_{q_2}^{\mu\nu}}
                            {q_2\cdot k}-
                        \frac{S_{p_1}^{\mu\nu}}
                            {p_1\cdot k}-
                        \frac{S_{p_2}^{\mu\nu}}
                            {p_2\cdot k}
                    \Big]
                \bigg\}\
                S_0^{(\mu)}+
                \mathcal{O}(\omega_k)
        \end{split}
        \end{equation}
        \item for $e^-(p_1)+\gamma(p_2)\rightarrow\gamma(q_1)+ e^-(q_2)$:
        \begin{equation}\label{eq:photon}
        \begin{split}
            S_{\text{tree}}^{(\gamma)}
                (\omega_k,\hat{k})=\!
                e \bigg\{
                    \Big[&
                        \frac{q_2\cdot
                                \epsilon_r^*(\vec{k})}
                            {q_2\cdot k}\!-\!
                        \frac{p_1\cdot
                                \epsilon_r^*(\vec{k})}
                            {p_1\cdot k}
                    \Big]\!+
                    i\epsilon^*_{r\mu}(\vec{k})
                    k_{\nu}
                    \Big[
                        \frac{\bar{J}_{q_2}^{\mu\nu}}
                            {q_2\cdot k}\!-\!
                        \frac{J_{p_1}^{\mu\nu}}
                            {p_1\cdot k}
                    \Big]
                \bigg\}\
                S_0^{(\gamma)}\!+\!
                \mathcal{O}(\omega_k)\!\!
        \end{split}
        \end{equation}
        \item for $e^-(p_1)+e^+(p_2)\rightarrow\gamma(q_1)+\gamma(q_2)$:
        \begin{equation}\label{eq:positron}
        \begin{split}
            S_{\text{tree}}^{(e)}
                (\omega_k,\hat{k})=\!
                e \bigg\{
                    \Big[&
                        \frac{p_2\cdot
                                \epsilon_r^*(\vec{k})}
                            {p_2\cdot k}\!-\!
                        \frac{p_1\cdot
                                \epsilon_r^*(\vec{k})}
                            {p_1\cdot k}
                    \Big]\!+
                    i\epsilon^*_{r\mu}(\vec{k})
                    k_{\nu}
                    \Big[
                        \frac{\bar{J}_{\bar{p}_2}^{\mu\nu}}
                            {p_2\cdot k}\!-\!
                        \frac{J_{p_1}^{\mu\nu}}
                            {p_1\cdot k}
                    \Big]
                \bigg\}\
                S_0^{(e)}\!+\!
                \mathcal{O}(\omega_k)\!\!
        \end{split}
        \end{equation}
    \end{itemize}
    The total angular momentum operators appearing in \eqref{eq:muon}, \eqref{eq:photon}, and \eqref{eq:positron} are given by the sum of the orbital and the spin components. For the incoming and the outgoing particles, these operators are given, respectively, by
    \begin{equation}
    \begin{split}
        J_p^{\mu\nu}=
            L_p^{\mu\nu}+
            S_p^{\mu\nu}
        , \hspace{5em}
        \bar{J}_q^{\mu\nu}=
            \bar{L}_q^{\mu\nu}+
            \bar{S}_q^{\mu\nu}
    \end{split}
    \end{equation}
    while for the incoming and the outgoing antiparticles, we have
    \begin{equation}
    \begin{split}
        \bar{J}_{\bar{p}}^{\mu\nu}=
            \bar{L}_{\bar{p}}^{\mu\nu}+
            \bar{S}_{\bar{p}}^{\mu\nu}
        \ \hspace{5em}
        J_{\bar{q}}^{\mu\nu}=
            L_{\bar{q}}^{\mu\nu}+
            S_{\bar{q}}^{\mu\nu}
    \end{split}
    \end{equation}
    The orbital components for incoming and outgoing particles satisfy 
    \begin{equation}\label{eq:orbital_particles}
    \begin{split}
        \bra{\Psi}
        L_p^{\mu\nu}
        \ket{p}=&
            +i\Big(
                p^{\mu}
                \frac{\partial}
                    {\partial p_{\nu}}-
                p^{\nu}
                \frac{\partial}
                    {\partial p_{\mu}}
            \Big)\
            \braket{\Psi|p}
        \\
        \bra{q}
        \bar{L}_q^{\mu\nu}
        \ket{\Psi}=&
            -i\Big(
                q^{\mu}
                \frac{\partial}
                    {\partial q_{\nu}}-
                q^{\nu}
                \frac{\partial}
                    {\partial q_{\mu}}
            \Big)\
            \braket{q|\Psi}
    \end{split}
    \end{equation}
    while for the antiparticles 
    \begin{equation}\label{eq:orbital_antiparticles}
    \begin{split}
        \bra{\Psi}
        \bar{L}_{\bar{p}}^{\mu\nu}
        \ket{\bar{p}}=&+i\Big(
                {p}^{\mu}
                \frac{\partial}
                    {\partial {p}_{\nu}}-
                {p}^{\nu}
                \frac{\partial}
                    {\partial {p}_{\mu}}
            \Big)\
            \braket{\Psi|\bar{p}}
        \\
        \bra{\bar{q}}
        L_{\bar{q}}^{\mu\nu}
        \ket{\Psi}=&
            -i\Big(
                {q}^{\mu}
                \frac{\partial}
                    {\partial {q}_{\nu}}-
                {q}^{\nu}
                \frac{\partial}
                    {\partial {q}_{\mu}}
            \Big)\
            \braket{\bar{q}|\Psi}
    \end{split}
    \end{equation}
    In \eqref{eq:orbital_particles} and \eqref{eq:orbital_antiparticles}, it is understood that the derivatives with respect to the hard particle momenta do not act on spinor wavefunctions. The corresponding spin components, on the other hand, act directly on the spinor wavefunctions and are given by \cite{Bern:2014vva}
    \begin{equation}\label{eq:spin_ops}
    \begin{split}
        \bra{\Psi}
        S_p^{\mu\nu}
        \ket{p}=&
        \frac{i}{4}[\gamma^{\mu},\gamma^{\nu}]\
            u(p)\
            \circ\
            \frac{\partial}
                {\partial u(p)}\
            \braket{\Psi|p}
        \\
        \bra{q}
        \bar{S}_q^{\mu\nu}
        \ket{\Psi}=&
        \bar{u}(q)\
            \frac{i}{4}[\gamma^{\mu},\gamma^{\nu}]\
            \circ\
            \frac{\partial}
                {\partial \bar{u}(q)}\
            \braket{q|\Psi}
    \end{split}
    \end{equation}
    for incoming and outgoing particles, and for antiparticles
    \begin{equation}\label{eq:spin_ops_anti}
    \begin{split}
        \bra{\Psi}
        \bar{S}_{\bar{p}}^{\mu\nu}
        \ket{\bar{p}}=&-
        \bar{\upsilon}(p)\
            \frac{i}{4}[\gamma^{\mu},\gamma^{\nu}]\
            \circ\
            \frac{\partial}
                {\partial \bar{\upsilon}(p)}\
            \braket{\Psi|\bar{p}}
        \\
        \bra{\bar{q}}
        S_{\bar{q}}^{\mu\nu}
        \ket{\Psi}=&-
        \frac{i}{4}[\gamma^{\mu},\gamma^{\nu}]\
            \upsilon(q)\
            \circ\
            \frac{\partial}
                {\partial \upsilon(q)}\
            \braket{\bar{q}|\Psi}
    \end{split}
    \end{equation}
    In \eqref{eq:spin_ops} and \eqref{eq:spin_ops_anti}, while we suppress spinor indices, we write the Dirac matrices and spinors in the appropriate order, ensuring correct index contraction. In addition, we use the symbol $\circ$ to indicate that the free spinor index of the commutator of the two gamma matrices is to be contracted with the spinor index carried by the derivative with respect to the fermionic wavefunction. In the definitions of the orbital and the spin components of the total angular momentum operators, $\ket{\Psi}$ stands for an arbitrary state, and 
    \begin{equation}\label{eq.3.2.9}
    \begin{split}
        \ket{q}=b^{\dagger}(\vec{q})\ket{0}
        ,\ \hspace{5em}
        \ket{p}=b^{\dagger}(\vec{p})\ket{0}
        \\
        \ket{\bar{q}}=d^{\dagger}(\vec{q})\ket{0}
        ,\ \hspace{5em}
        \ket{\bar{p}}=d^{\dagger}(\vec{p})\ket{0}
    \end{split}
    \end{equation}
    are single particle/antiparticle fermionic states. 

\subsection{Applying subleading dressings to Fock states}
    We define final and initial dressed states, for the three processes we consider, as follows \cite{Kulish:1970ut,Choi:2019rlz,Tomaras_2020}
    \begin{equation}\label{eq:sub_FK_states}
    \begin{split}
        \ket{q_1,q_2}_d=
            e^{
                \int\reallywidetilde{d^3k}\
                [f_q(\vec{k})
                    \cdot 
                a^{\dagger}(\vec{k})
                -h.c.]
            }
            \Big(
                1+
                \int\reallywidetilde{d^3k}\
                [g_q(\vec{k})
                    \cdot
                a^{\dagger}(\vec{k})
                -h.c.]
                +\mathcal{O}(e^2)
            \Big)
            \ket{q_1,q_2}\ 
        \\ 
        \ket{p_1,p_2}_d=
            e^{
                \int\reallywidetilde{d^3k}\
                [f_p(\vec{k})
                    \cdot 
                a^{\dagger}(\vec{k})
                -h.c.]
            }
            \Big(
                1+
                \int\reallywidetilde{d^3k}\
                [g_p(\vec{k})
                    \cdot
                a^{\dagger}(\vec{k})
                -h.c.]
                +\mathcal{O}(e^2)
            \Big)
            \ket{p_1,p_2}
    \end{split}
    \end{equation}
    where $\ket{q_1,q_2}, \ket{p_1,p_2}$ are two-particle Fock states, which are taken to be hard. That is, their energies are greater than $E_d$, the infrared energy scale characterizing the soft photons in the clouds. Here $f$ and $g$ denote the leading and subleading dressing functions. As in \eqref{eq:dot_prod_leading}, the dot product between the subleading dressing functions and the photon creation operator is given by
    \begin{equation}\label{eq.3.2.3}
        g_q(\vec{k})
            \cdot 
        a^{\dagger}(\vec{k})=
            g_q^{\mu}
                (\vec{k})
            \sum_{r}
                \epsilon_{r\mu}^*
                    (\vec{k})
                a_r^{\dagger}
                    (\vec{k}),
        \hspace{3em}
        g_p(\vec{k})
            \cdot 
        a^{\dagger}(\vec{k})=
            g_p^{\mu}
                (\vec{k})
            \sum_{r}
                \epsilon_{r\mu}^*
                    (\vec{k})
                a_r^{\dagger}
                    (\vec{k})
    \end{equation}
    and similarly for the dot product of the complex conjugate of the dressing functions with the annihilation operator. The subscript $q$ in the dressing functions indicates dependence on both the momenta of the hard outgoing particles, while $p$ indicates dependence on both the momenta of the hard incoming particles. 
    
    The leading dressing functions for the outgoing and incoming states are explicitly given by \cite{Kulish:1970ut, Tomaras_2020}:
    \begin{itemize}
        \item for $e^-(p_1)+\mu^-(p_2)\rightarrow e^-(q_1)+\mu^-(q_2)$:
        \begin{equation}\label{eq:dressing_fncs_mu}
        \begin{split}
            f_q^{\mu}(\vec{k})&=
                e\bigg[
                    e^
                        {-iq_1\cdot kt_0/q_1^0}
                    \Big(
                        \frac{q_1^{\mu}}
                            {q_1\cdot k}
                        -c^{\mu}
                    \Big)~+
                    e^
                        {-iq_2\cdot kt_0/q_2^0}
                    \Big(
                        \frac{q_2^{\mu}}
                            {q_2\cdot k}
                        -c^{\mu}
                    \Big)
                \bigg]
            \\
            f_p^{\mu}(\vec{k})&=
                e\bigg[
                    e^
                        {-ip_1\cdot kt_0/p_1^0}
                    \Big(
                        \frac{p_1^{\mu}}
                            {p_1\cdot k}
                        -c^{\mu}
                    \Big)+
                    e^
                        {-ip_2\cdot kt_0/p_2^0}
                    \Big(
                        \frac{p_2^{\mu}}
                            {p_2\cdot k}
                        -c^{\mu}
                    \Big)
                \bigg]
        \end{split}
        \end{equation}
        \item for $e^-(p_1)+\gamma(p_2)\rightarrow\gamma(q_1)+ e^-(q_2)$:
        \begin{equation}\label{eq:dressing_fncs_gamma}
        \begin{split}
            f_q^{\mu}(\vec{k})&=
                e\
                    e^
                        {-iq_2\cdot kt_0/q_2^0}
                    \Big(
                        \frac{q_2^{\mu}}
                            {q_2\cdot k}
                        -c^{\mu}
                    \Big)
            \\
            f_p^{\mu}(\vec{k})&=
                e\
                    e^
                        {-ip_1\cdot kt_0/p_1^0}
                    \Big(
                        \frac{p_1^{\mu}}
                            {p_1\cdot k}
                        -c^{\mu}
                    \Big)
        \end{split}
        \end{equation}
        \item for $e^-(p_1)+e^+(p_2)\rightarrow\gamma(q_1)+\gamma(q_2)$:
        \begin{equation}\label{eq:dressing_fncs_e}
        \begin{split}
            f_q^{\mu}(\vec{k})&=
                0
            \\
            f_p^{\mu}(\vec{k})&=
                e\bigg[
                    e^
                        {-ip_1\cdot kt_0/p_1^0}
                    \Big(
                        \frac{p_1^{\mu}}
                            {p_1\cdot k}
                        -c^{\mu}
                    \Big)-
                    e^
                        {-ip_2\cdot kt_0/p_2^0}
                    \Big(
                        \frac{p_2^{\mu}}
                            {p_2\cdot k}
                        -c^{\mu}
                    \Big)
                \bigg]
        \end{split}
        \end{equation}
    \end{itemize}
    The subleading dressing functions are determined by implementing the prescription in \cite{Choi:2019rlz}:
    \begin{itemize}
        \item for $e^-(p_1)+\mu^-(p_2)\rightarrow e^-(q_1)+\mu^-(q_2)$:
        \begin{equation}\label{eq:dressing_fncs_mu_sub}
        \begin{split}
            g_q^{\mu}(\vec{k})&=
                ie k_{\nu}
                \bigg[
                    e^
                        {-iq_1\cdot kt_0/q_1^0}
                    \frac{\bar{J}_{q_1}^{\mu\nu}}
                        {q_1\cdot k}~+
                    e^
                        {-iq_2\cdot kt_0/q_2^0}
                    \frac{\bar{J}_{q_2}^{\mu\nu}}
                        {q_2\cdot k}
                \bigg]\\
            g_p^{\mu}(\vec{k})&=
                ie k_{\nu}
                \bigg[
                    e^
                        {-ip_1\cdot kt_0/p_1^0}
                    \frac{J_{p_1}^{\mu\nu}}
                        {p_1\cdot k}+
                    e^
                        {-ip_2\cdot kt_0/p_2^0}
                    \frac{J_{p_2}^{\mu\nu}}
                        {p_2\cdot k}
                \bigg]
        \end{split}
        \end{equation}
        \item for $e^-(p_1)+\gamma(p_2)\rightarrow\gamma(q_1)+ e^-(q_2)$:
        \begin{equation}\label{eq:dressing_fncs_gamma_sub}
        \begin{split}
            g_q^{\mu}(\vec{k})=
                ie k_{\nu}\
                    e^
                        {-iq_2\cdot kt_0/q_2^0}
                    \frac{\bar{J}_{q_2}^{\mu\nu}}
                        {q_2\cdot k}\\
            g_p^{\mu}(\vec{k})=
                ie k_{\nu}\
                    e^
                        {-ip_1\cdot kt_0/p_1^0}
                    \frac{J_{p_1}^{\mu\nu}}
                        {p_1\cdot k}
        \end{split}
        \end{equation}
        \item for $e^-(p_1)+e^+(p_2)\rightarrow\gamma(q_1)+\gamma(q_2)$:
        \begin{equation}\label{eq:dressing_fncs_e_sub}
        \begin{split}
            g_q^{\mu}(\vec{k})&=
                0
            \\
            g_p^{\mu}(\vec{k})&=
                ie k_{\nu}
                \bigg[
                    e^
                        {-ip_1\cdot kt_0/p_1^0}
                    \frac{J_{p_1}^{\mu\nu}}
                        {p_1\cdot k}-
                    e^
                        {-ip_2\cdot kt_0/p_2^0}
                    \frac{\bar{J}_{\bar{p}_2}^{\mu\nu}}
                        {p_2\cdot k}
                \bigg]
        \end{split}
        \end{equation}
    \end{itemize}
    The integrands in \eqref{eq:sub_FK_states} are dominated by low momenta. We are, therefore, justified in taking $E_d << 1/t_0$, thereby approximating the phases $e^{-iq_1\cdot k t_0/q_1^0}$, etc, appearing
    in \eqref{eq:dressing_fncs_mu} to \eqref{eq:dressing_fncs_e_sub} by unity.
    
\subsection{Dressed amplitudes: infrared-finiteness and soft-photon suppression}
    Using the dressings in \eqref{eq:sub_FK_states}, the dressed elastic amplitudes for the processes under consideration can be shown to be infrared-finite, order by order in perturbation theory. In addition, taking the infrared energy scale $\Lambda$ to be equal to the dressing scale $E_d$, one can infer that the dressed elastic amplitudes for these three processes are equal to the infrared-finite part of their undressed counterparts, up to negligible power-law corrections of the dressing scale $E_d$.
    
\begin{figure}
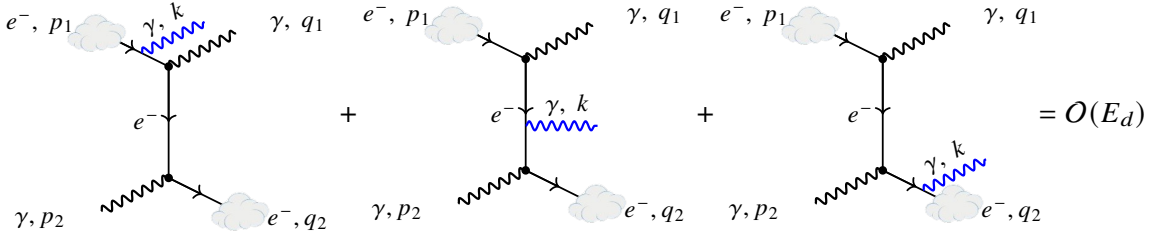

        \centering
        \begin{minipage}[c]{\linewidth}
        \vspace{0.5em}
        \centering
            \begin{minipage}[c]{0.3\linewidth}
                \centering
                \resizebox{\linewidth}{!}{\input{Sections/Figure_4a}}
            \end{minipage}
            $\!\!+\!\!$
            \begin{minipage}[c]{0.3\linewidth}
                \centering
                \resizebox{\linewidth}{!}{\input{Sections/Figure_4b}}
            \end{minipage}
            $\!\!+\!\!$
            \begin{minipage}[c]{0.3\linewidth}
                \centering
                \resizebox{\linewidth}{!}{\input{Sections/Figure_4c}}
            \end{minipage}
            $\!\!\!\!\!=\mathcal{O}(E_d)$
        \end{minipage}
        \caption{
                Tree-level Feynman diagrams depicting soft-photon emission during the scattering of a dressed electron with a hard photon. The overall contributions from these diagrams to the scattering amplitude are suppressed. Only the u-channel diagrams are shown in the figure. The corresponding s-channel diagrams are obtained by exchanging the two final state particle legs, and their overall contributions are likewise independently suppressed. The emitted soft photons that do not belong in the dressing clouds are drawn in blue.
        }
        \label{fig:dressed_suppression}
    \end{figure}

    We now focus on processes involving the emission of an additional soft photon with momentum $\vec{k}$ and energy $\omega_k=|\vec{k}|<E_d$, and argue that the dressed amplitude for this emission is highly suppressed, at least to leading order in the QED coupling constant. Indeed, the single-photon-emitting dressed amplitude, defined as
    \begin{equation}
        \tilde{S}^{(\alpha)}=
            \tensor[_{d}]{\braket{q_1,q_2;k|
                S|
            p_1,p_2}}{_{d}},
            \hspace{2em}
            \alpha\in\{\gamma,\ \mu,\ e\}
    \end{equation}
    can be shown to vanish in the limit $k\to 0$ at tree level \cite{Christodoulou:2025jus}:
    \begin{equation}
        \Tilde{S}_{\text{tree}}^{(\alpha)}=
            \mathcal{O}(E_d)
    \end{equation}
    This indicates that the subleading corrections of the dressing functions completely suppress the emission of soft photons in the deep infrared. The fact that such soft radiation does not appear in the asymptotic outgoing Hilbert space implies that the cross sections obtained using the Bloch-Nordsieck method can be reproduced using dressed amplitudes \cite{Choi:2019rlz}. \figref{fig:dressed_suppression} illustrates the suppression of soft-photon emission in the dressed state formalism at tree-level for Compton scattering. Analogous diagrams can be drawn for the electron scattered off by the heavier muon and for the electron-positron annihilation into two photons.

\section{Summary and future directions}
    \label{sec:sum}
    Faddeev-Kulish states, constructed by dressing charged particles with coherent clouds of soft photons, capture the long-range dynamics of gauge and gravitational interactions at large asymptotic times. Consequently, these states, in contrast to bare Fock states that contain zero or a finite number of soft photons, provide the appropriate basis for defining the S-matrix elements. One hopes that the dressed S-matrix will be free from any infrared divergences to all orders in perturbation theory.
In addition, if extended to subleading order in the soft expansion, the Faddeev-Kulish dressings also suppress soft-photon emission with energies less than the characteristic scale $E_d$, at least at tree level. Thus, the dressed state formalism is equivalent to the Bloch–Nordsieck method.

It is not clear whether this suppression of soft radiation in the deep infrared can be trusted at higher orders in perturbation theory. 
Indeed, at higher orders, the subleading soft theorem receives logarithmic corrections in the energy of the emitted soft photon, raising the question of how to construct loop-corrected Faddeev-Kulish dressings that consistently account for these effects. An important direction for future work is to construct such dressings by systematically analyzing soft-photon-emitting amplitudes at the loop level. These loop-corrected dressings are expected to remove the additional logarithmic divergences that arise at higher orders in perturbation theory \cite{Bern:2014oka,He:2014bga,Mao:2017wvx,Sahoo:2018lxl,Laddha:2018myi,Saha:2019tub,Sahoo:2019yod,Sahoo:2020ryf,Delisle:2020uui,Krishna:2023fxg}.

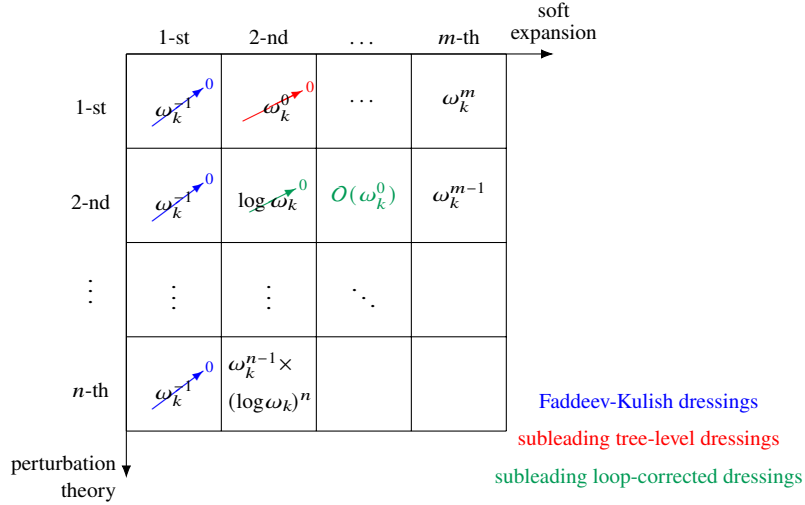
\begin{figure}[!t]
    \centering
        \begin{tikzpicture}
        [>=Latex, 
        scale=1.25, 
        every node/.style={font=\scriptsize}]

    \path[use as bounding box] (-0.1,-0.25) rectangle (4.1,4.25);
    \draw[step=1] (0,0) grid (4,4);
    
    \draw[->] (0,4) -- (0,-0.5);

    \node[left] at (0,-0.5)
        {\shortstack[r]{perturbation\\ theory}};
    \node[above] at (4.5,4)
        {\shortstack[c]{soft\\ expansion}};
    \node[above] at (0.5,4){$1$-st};
    \node[above] at (1.5,4){$2$-nd};
    \node[above] at (2.5,4){\dots};
    \node[above] at (3.5,4){$m$-th};
    \node[above] at (-0.375,3.25){$1$-st};
    \node[above] at (-0.375,2.25){$2$-nd};
    \node[above] at (-0.375,1.25){\vdots};
    \node[above] at (-0.375,0.25){$n$-th};
    
    \draw[->] (0,4) -- (4.5,4);

    \node at (0.5,3.5) 
        {$\CancelTo[\color{blue}]{0}{\omega_k^{-1}}$};
    \node at (0.5,2.5) 
        {$\CancelTo[\color{blue}]{0}{\omega_k^{-1}}$};
    \node at (0.5,1.5) {\vdots};
    \node at (0.5,0.5) 
        {$\CancelTo[\color{blue}]{0}{\omega_k^{-1}}$};

    \node at (1.5,3.5) 
        {$\CancelTo[\color{red}]{0}{\hspace{0.75em}\omega_k^0}$};
    \node at (1.5,2.5) 
        {$\CancelTo[\color{ForestGreen}]{0}
            {\!\!\!\!\log\omega_k\!\!\!\!}$};
    \node at (1.5,1.5) {\vdots};
    \node at (1.5,0.7) {$\!\!\!\!\!\omega_k^{n-1}\times$};
    \node at (1.5,0.3) {$\ \!(\log\!\omega_k\!)^n$};

    \node at (2.5,3.5) {\dots};
    \node at (2.5,2.5) {${\color{ForestGreen}\mathcal{O}(\omega_k^0)}$};
    \node at (2.5,1.5) {$\ddots$};
    \node at (2.5,0.5) {};
    
    \node at (3.5,3.5) {$\omega_k^m$};
    \node at (3.5,2.5) {$\omega_k^{m-1}$};
    \node at (3.5,1.5) {};
    \node at (3.5,0.5) {};

    \node at (5.5,0.3) {\color{blue} Faddeev-Kulish dressings};
    \node at (5.5,-0.1) {\color{red} subleading tree-level dressings};
    \node at (5.5,-0.5) {\color{ForestGreen} subleading loop-corrected dressings};
    
    \end{tikzpicture}
    \caption{The horizontal axis of this figure represents the orders of magnitude in the soft-momentum expansion, while the vertical axis denotes the order in perturbation theory. At subleading order in the soft expansion, higher loop orders introduce additional powers of $\omega_k$ multiplying the powers of the divergent logarithms. In the soft limit, these subleading contributions at higher orders in perturbation theory vanish, rendering higher-loop corrections, beyond the one-loop case, to the dressings unnecessary.}
    \label{fig:rad_emitting_dependence_table}
\end{figure}

Even if loop-corrected subleading divergences are removed, soft-photon-emitting amplitudes are still expected to develop constant, non-vanishing contributions in the limit where the emitted soft-photon energy tends to zero. 
As a result, 
extending loop-corrected subleading dressings to next-to-subleading order in the soft expansion may be necessary in order to guarantee the vanishing of such amplitudes. 
However, regardless of whether such extensions are implemented, the loop-corrected subleading dressings should suffice to render soft-photon-emitting amplitudes finite to all orders in perturbation theory, since higher loop orders introduce additional powers of the soft energy multiplying the subleading logarithmic divergences \cite{Krishna:2023fxg}, as illustrated in \figref{fig:rad_emitting_dependence_table}.

Another promising next step is to investigate whether analogous subleading dressings can be systematically constructed in gravitational scattering \cite{He:2014laa,Cachazo:2014fwa,Strominger:2014pwa,Banks:2014iha,Strominger:2013jfa,Chakrabarti:2017zmh,Beneke:2021umj}, thereby providing explicit realizations of the gravitational S-matrix beyond the leading soft approximation. Natural processes to explore in this direction include soft-graviton emission in matter scattering, as well as pure graviton $2\rightarrow2$ amplitudes at tree level. These examples offer concrete testing grounds for the role of soft gravitons in rendering the S-matrix infrared finite, and for exploring how correlations between soft and hard sectors manifest in gravity \cite{Carney:2017oxp,Mirbabayi:2016axw,Carney:2017jut,Toumbas:2023qbo,Hawking:2016msc,Gabai:2016kuf,Grignani:2016igg,Irakleous:2021ggq}. Carrying out such explicit computations may clarify whether gravitational subleading dressings can play a role analogous to their QED counterparts.
\section*{Acknowledgements}
    This work was partially supported by the Cyprus Research and Innovation Foundation grant EXCELLENCE/0421/0362. 

\bibliography{References/References_1,References/References_2,References/References_3,References/References_4}
    \bibliographystyle{jhep}


\end{document}